\documentstyle[axodraw,12pt]{article}

\begin{document}
\def\bra{\langle}
\def\ket{\rangle}

{\Large {Weak radiative hyperon decays in quark model }}

\vskip 10mm
{\large{ E.N.Dubovik$^\dagger$\footnote{dubovik@thsun1.jinr.ru}, 
V.S.Zamiralov$^*$\footnote{zamir@depni.sinp.msu.ru}, S.N.Lepshokov$^*$}}
\vskip 5mm
$^\dagger$ Bogoliubov Laboratory of Theoretical Physics, JINR (Dubna)
\vskip 3mm
$^*$ D.V.Skobeltsyn Institute of Nuclear Physics,
Moscow State University, Moscow
\vskip 10mm

Weak radiative hyperon decays are considered in the framework of
a quark model. The phenomenological model includes 1-quark transitions
with the effective $sd\gamma$-vertex and 2-quark ones with the W-exchange
$s+u\rightarrow u+d+\gamma$ and turns out to describe well
the data and gives predictions for asymmetry of the decays
$\Lambda\rightarrow n+\gamma$ and  $\Xi^-\rightarrow \Sigma^- +\gamma$.
\vskip 10mm
\section{Introduction}

Recently, some new experiments have been performed which measured
widths and asymmetry parameters in weak radiative hyperon decays
(WRHD). The situation changed significantly for the decays
$\Xi^{0} \rightarrow \Lambda \gamma$  and $\Xi^{0} \rightarrow
\Sigma^{0} \gamma$, where asymmetry parameters turned out to be
negative \cite{PDG}, while previous experiments indicated positive
asymmetry \cite{PDGold}. In this connection, interest was revived
in the old problem of describing of WRHD's either with the Hara
theorem \cite{Hara} stating zero asymmetry in $\Sigma^{+}
\rightarrow p \gamma$ decay, or without it.

In the quark model WRHD's are described by three kinds of diagrams
(see, e.g., \cite{Sharma}), namely, 1-quark diagrams with photon
emission from the effective $sd\gamma$ -- vertex, 2-quark diagrams
with bremstrahlung and $W$-exchange and the 3rd quark as a
spectator, and, finally, diagrams where 2 quarks exchange
$W$-boson while the 3rd quark emitts photon. As a rule, the latter
class of diagrams can be safely neglected \cite{Verma}.

However, 1-quark diagram contributions are not able to explain the
observed radiative rates. Even contributions  of the penguin-like
diagrams are not strong enough to enhance sufficiently the $s
\rightarrow d + \gamma$ decay rate. At the same time, as it was
shown in ~\cite{Sharma}-\cite{Ryaz}, 2-quark diagram proved to be
important. However, calculations of these contributions in the
quark model simultaneously for parity conserving and parity
violating amplitudes without phenomenological parameters do not
yield agreement with the data, in particular with taking account
of the new experiments \cite{PDG}.

That is why parity violating amplitudes are often calculated
within the quark models of type \cite{Verma}, whereas parity
conserving ones are treated with  the help of the unitary models
of nonleptonic decays, vector-dominance hypothesis and $SU(6)_W$
symmetry \cite{ZenM}, \cite{Zen}, \cite{Zen1}. Recently
\cite{Zen06}, it was attempted to describe in a unique way weak
nonleptonic and radiative hyperon decays. Taking account of the
complexity of the problem, the author mainly succeeded in
describing of the radiative hyperon decays.

We would like to propose on the basis of the quark model a
phenomenological model which however opens the way to theoretical
analysis of the problem. We also show that quark models like those
of \cite{Sharma} are related to models based on unitary symmetry
and pole models \cite{Zen}, \cite{Zen1}, \cite{Zen06}. Numerical
analysis partly follow the lines of \cite{Neu}.

\section{Kinematics of the weak  radiative hyperon decay}

A gauge-invariant form of the amplitude of the weak radiative
decay $B_i\rightarrow B_f+\gamma$ is usually written in the
following way:
\begin{equation}
{\cal{A}}_W=
\bar B_f (B^{PC}+A^{PV}\gamma_5)\hat k {\hat \epsilon} B_i,
\label{decay}
\end{equation}
where $B_i$ и $B_f$ are Dirac spinors of the initial and final
baryons, $q$ being 4-momentum of the photon ($k_{c.m.}\equiv
k_\gamma$ is respectively the 3-momentum of the photon in the rest
frame of the initial baryon), $\epsilon_\mu$ being photon
polarization 4-vector.

The partial width of the radiative hyperon decay in terms of the
phenomenological parity-violating (PV) $A^{PV}$ and
parity-conserving (PC)  $B^{PC}$ amplitudes are given
by~\cite{Behr},\cite{Pak}
\begin{equation}
\Gamma_\gamma=\frac{k_{c.m.}^3}{\pi}(|A^{PV}|^2+|B^{PC}|^2),
\label{rate}
\end{equation}
while the corresponding asymmetry is written as
\begin{equation}
A_\gamma=\frac{2Re(A^{PV *}B^{PC})}{|A^{PV}|^2+|B^{PC}|^2}.
\label{alpha}
\end{equation}

Experimental data on rates and branching ratios
$BR = \Gamma_\gamma / \Gamma(\rm total)$ and asymmetry parameters
\cite{PDG} are given in Table 1.

\section{On the 1-quark amplitudes of the radiative hyperon decay }

Amplitudes of the radiative hyperon decay due to 1-quark
transition $ s \rightarrow d + \gamma $ were calculated in plenty
of works, and we put the main results into the 2nd column of Table
2.

Already from the partial decay widths it could be seen that
1-quark amplitudes should give a small contribution to the real
partial widths. This can be stated, e.g., from the data on $
\Xi^{-,0} \rightarrow \Sigma^{-,0}\gamma$ decays. Really, the $
\Xi^{-} \rightarrow \Sigma^{-} \gamma$ decay can be explained
(within the quark model) only by the 1-quark diagram, and its
reduced partial width  in the units  $10^{-7}\mu_N$ is about 4.0 (
see the 4th column of the Table 5), whereas the reduced partial
width of the $ \Xi^{0} \rightarrow \Sigma^{0} \gamma$ decay, equal
to it in the model with the only 1-quark diagram, turns out to be
$\sim$60.0 in the same units, which show the necessity to go out
of the 1-quark diagram description.

At the same time, 1-quark diagram contributions to other radiative
decays, including $ \Sigma^{+} \rightarrow p \gamma $ one, turn
out to be even more suppressed, as it can be seen from Table 2.

Even an enhancement due to the penguin diagrams does not solve the
problem \cite{Penguin}.

Note that the standard  1-quark diagram contribution is given by
the effective weak strangeness-violating neutral current
\begin{eqnarray}
{\cal{J}}_\mu^W= (D_{ds}+F_{ds})\bar B^{\alpha}_{2}{\cal{O}}_
\mu B^{3}_{\alpha}+
(D_{ds}-F_{ds})\bar B_{\alpha}^{3}{\cal{O}}_\mu B_{2}^{\alpha},
\label{eff}
\end{eqnarray}
where $B^{\alpha}_{\beta}$ is the baryon octet, $\alpha,\beta
=1,2,3$, и $B^{3}_{1}=p$ etc. Really, putting $D_{ds}=-b$,
$F_{ds}/D_{ds}=2/3$ one reproduces standard quark model results
(see 2nd column of Table 2 and, e.g., \cite{Sharma}), and in
applications, generally speaking, one takes $b^{PC}\not=b^{PV}$ (
the same for $D_{ds},F_{ds}$).

\section{On the structure of the 2-quark amplitudes
}

Let us consider now contributions of the 2-quark weak
radiative transitions $ s+u \rightarrow u+d + \gamma $
with $W$-exchange. ( They were calculated,for example, in
\cite{Sharma}, \cite{Verma}, \cite{Ryaz}. )

We begin with the analysis of a set of diagrams of the $
\Sigma^{+} \rightarrow p \gamma $--decay. ( For other decays see
Appendix A.) The matrix element of the $ \Sigma^{+} \rightarrow p
\gamma $ decay can be put in the form
\begin{eqnarray}
6<p_{\downarrow}, \gamma(+1) |O| \Sigma^{+}_{\uparrow} >=
\\
=<2u_{2}u_{2}d_{1}-u_{2}d_{2}u_{1}-
d_{2}u_{2}u_{1},\gamma(+1)|O|2u_{1}u_{1}s_{2}-u_{1}s_{1}u_{2}-
s_{1}u_{1}u_{2}>=
\nonumber\\
=4<u_{2}u_{2}d_{1},\gamma(+1) |O|u_{1}u_{1}s_{2}>-
4<u_{2}u_{2}d_{1},\gamma(+1) |O|u_{1}s_{1}u_{2}>-
\nonumber\\
-4<u_{2}d_{2}u_{1},\gamma(+1) |O|u_{1}u_{1}s_{2}>+
4<u_{2}d_{2}u_{1},\gamma(+1) |O|u_{1}s_{1}u_{2}>, \quad
\nonumber
\label{sigma}
\end{eqnarray}
where $ q_{1,2}$ means states with definite spin projection $
q_{\uparrow,\downarrow} $ of the quark inside the baryon. An
explicit form of the operator $O=O^{PV}+O^{PC}$ is for a moment
irrelevant to us. The 1st matrix element (m.e.) in the right-hand
side (RHS) of the last expression, Eq.(\ref{sigma}), $
<u_{2}u_{2}d_{1},\gamma(+1) |O|u_{1}u_{1}s_{2}>$ in the case of
the $W$-exchange between the quarks is described by the 1st
diagram of Fig.1, because this m.e. cannot be represented by a
diagram with a spectator. It is plausible to assume that its
contribution is small (see, e.g., \cite{Verma}).

There are three different diagrams of Fig.2 with the  $ u_{2}$
quark as a spectator which give a contribution to the 2nd m.e. of
the RHS of Eq.(\ref{sigma}) $ <u_{2}u_{2}d_{1},\gamma(+1)
|O|u_{1}s_{1}u_{2}>=A_{1} $, and
$$ A_{1}=e_u^f A+e_s^i E+e_u^i B
=\frac{2}{3}A-\frac{1}{3}E+\frac{2}{3}B. $$ Here $A$ corresponds
to the non-flip transition quark amplitude with all quarks having
spin projection +1/2: $ s_{\uparrow}+u_{\uparrow} \rightarrow
u_{\uparrow}+d_{\uparrow} $; $E$ corresponds to the non-flip
transition quark amplitude with quarks having different spin
projections: $ s_{\downarrow}+u_{\uparrow} \rightarrow
u_{\downarrow}+ d_{\uparrow} $ while $B$ corresponds to the
spin-flip transition quark amplitude with $ s$ quark spin
projection equal to +1/2 $ s_{\uparrow}+u_{\downarrow} \rightarrow
u_{\downarrow}+d_{\uparrow} $; numbers of the coefficients are
just electric charges of quarks.

There are three different diagrams of Fig.2 with the  $
 u_{1}$ quark as a spectator which give a contribution
to the 3rd m.e. of the RHS of Eq.(\ref{sigma}) $
<u_{2}d_{2}u_{1},\gamma(+1) |O|u_{1}u_{1}s_{2}>=A_{3}$:
$$ A_{3}=e_u^f C+e_d^f E+e_u^i \tilde A=
\frac{2}{3}C-\frac{1}{3}E+\frac{2}{3} \tilde A, $$ where two new
coefficients are introduced: $C$ corresponds to the spin-flip
transition quark amplitude with $ s$ quark spin projection equal
to -1/2 $ s_{\downarrow}+u_{\uparrow} \rightarrow
u_{\uparrow}+d_{\downarrow} $, while $ \tilde A$ corresponds to
the non-flip transition quark amplitude with all quarks having
spin projection -1/2: $ s_\downarrow+u_\downarrow \rightarrow
u_\downarrow+d_\downarrow.$ To the 4th m.e. of the RHS of
Eq.(\ref{sigma}) $ <u_{2}d_{2}u_{1},\gamma(+1)
|O|u_{1}s_{1}u_{2}>$ there are two kinds of contributions, the
first given by the diagrams of Fig.4 with quark $ u_{2}$ as a
spectator
$$ A_{2}=e_d^f A+e_s^i C+e_u^i D=
-\frac{1}{3}A-\frac{1}{3}C+\frac{2}{3}D, $$ where $D$ corresponds
to the non-flip transition quark amplitude with different spin
projections of quarks, $ s_{\uparrow}+u_{\downarrow} \rightarrow
u_{\uparrow}+d_{\downarrow} $, and, the second given by the
diagrams of Fig.4 with quark $ u_{1}$ as a spectator:
$$ A_{4}=e_s^i \tilde A+e_d^f B+e_u^f D=
\frac{2}{3}D-\frac{1}{3} \tilde A-\frac{1}{3}B. $$

Totally there are 4 combinations of the m.e.'s $A, \tilde A, B, C,
D, E $ denoted as $ A_{k},\, k=1, 2, 3, 4.$ Finally, we have
\begin{equation}
<p_{\downarrow}, \gamma(+1) |O| \Sigma^{+}_{\uparrow} >=
\frac{2}{3}(-2A_{1}+A_{2}-2A_{3}+A_{4}).
\end{equation}

In a similar way one can obtain expressions for the amplitudes of
all the other radiative decays.

Upon assuming that spectator quarks do not change amplitudes $
A_{k},\, k=1, 2, 3, 4,$ while going from one decay to another all
the amplitudes can be expressed in terms of these quantities
\begin{eqnarray}
<p_{\downarrow}, \gamma(+1) |O| \Sigma^{+}_{\uparrow} >=
\frac{2}{3}(-2A_{1}+A_{2}-2A_{3}+A_{4}),
\label{AAA} \\
<n_{\downarrow}, \gamma(+1) |O| \Sigma^{0}_{\uparrow} >=
\frac{2}{3\sqrt{2}}(A_{1}-2A_{2}-2A_{3}+A_{4}),
\nonumber\\
<n_{\downarrow}, \gamma(+1) |O| \Lambda_{\uparrow} >=
\frac{2}{\sqrt{6}}(A_{1}-2A_{2}-A_{4}),
\nonumber\\
<\Lambda_{\downarrow}, \gamma(+1) |O| \Xi^{0}_{\uparrow} >=
\frac{2}{\sqrt{6}}(A_{1}-A_{2}),
\nonumber\\
<\Sigma^{0}_{\downarrow}, \gamma(+1) |O| \Xi^{0}_{\uparrow} >=
\frac{2}{3 \sqrt{2}}(A_{1}+A_{2}-2A_{3}+4A_{4}).
\nonumber
\end{eqnarray}
The 2-quark amplitudes satisfy the following relations:
$$
<p_{\downarrow}, \gamma(+1) |O| \Sigma^{+}_{\uparrow} >+
2\sqrt{6}<\Lambda_{\downarrow}, \gamma(+1) |O| \Xi^{0}_{\uparrow} >=
$$
$$ \sqrt{2}<\Sigma^{0}_{\downarrow}, \gamma(+1) |O| \Xi^{0}_{\uparrow} >+
\sqrt{6}<n_{\downarrow}, \gamma(+1) |O| \Lambda_{\uparrow} >,
$$
$$
\sqrt{2}<n_{\downarrow}, \gamma(+1) |O| \Sigma^{0}_{\uparrow} >=
<p_{\downarrow}, \gamma(+1) |O| \Sigma^{+}_{\uparrow} >+
\sqrt{6}<\Lambda_{\downarrow}, \gamma(+1) |O| \Xi^{0}_{\uparrow} >,
$$
as they depend not on all $ A_{k}$s, but only on their linear
combinations $ B_{1} =A_{1}- A_{2},\,  B_{2}=A_{2}+ A_{3},\,
B_{31}=A_{3}- A_{4}$. It is straightforward to show that in the
model of \cite{Sharma}  $ A_{k}$s are given by
\begin{equation} \begin{array}{l}
A_{1}^{PV}=A_{1}^{PC}=\frac{1}{6}(1-2\zeta)(1+X),\quad
A_{2}^{PV}=A_{2}^{PC}=-\frac{1}{6}(1-2\zeta)(1-X),
\\
A_{3}^{PV}=A_{4}^{PC}=\frac{1}{6}(1+X),\qquad \quad
A_{4}^{PV}=A_{3}^{PC}=-\frac{1}{6}(1+X),
\nonumber
\end{array}
\end{equation}
where $ X=k/2m_{u}$ and $ 6\zeta=(1-m_{u}/m_{s})=(1-\epsilon) $
\cite{Sharma}. One can see that up to a factor this formula
reproduces their results (see Table 1 in \cite{Sharma} ) and the
3rd column in Table 4 of the present work.

Calculations of the quark diagrams along the lines of
\cite{Sharma} allow one to find also the amplitudes $A, \tilde A,
B,C,D,E$ in the first order in $k$ and linear symmetry breaking by
the mass of the strange quark $m_s$. (The last assertion means
that if a photon is emitted by the strange quark of the
corresponding amplitudes $A, \tilde A, B,C,D,E$, one should put a
factor $\epsilon=m_u/m_s, m_u=m_d$. Later we will see that this
factor in fact plays the role of one more free parameter taking
either positive or negative values.)

Namely, in this approximation and in the units of $(eG_F $
$sin\theta_C cos\theta_C /\sqrt{2}m_u)$ $|\psi(k)|^2$ c $X=k/2m_u$
(see \cite{Sharma}) the PC-amplitudes have the form

$A^{PC}=\tilde A^{PC}=0$,
$B^{PC}=-(1+X)$, $C^{PC}=-(1+X)$,

$D^{PC}=E^{PC}=1+X$,

and these expressions are valid for the photon emitted off the
quark with the spin projection +1/2. When a photon is emitted off
the quark with  the spin projection -1/2 one should change the
sign of $X$ in the amplitudes $D^{PC}$ and $C^{PC}$:

$D^{PC}_{-1/2}=1-X$, $C^{PC}_{-1/2}=-(1-X)$.

Instead for the PV-amplitudes the relations

$A^{PV}=\tilde A^{PV}=0$,
$B^{PV}=(1+X)$, $C^{PV}=(1-X)$, $D^{PV}=1+X$, $E^{PV}=-(1+X)$

are valid for both spin projections of the emitted quark.

Putting $ X=0,\, \epsilon=1-6\zeta$ in the PV-amplitudes, one
get's the results of \cite{Zen} for the 2-quark transitions with
the 3rd quark as a spectator (dividing by $\sqrt{2}$ in order to
obtain exactly the coefficients of $b$ in Eq.(5.2a) of
\cite{Zen}). The results of \cite{Sharma} are put in the 3rd
columns of Tables 3,4, where the overall factor

$(\kappa_0/m_u/\sqrt{2})G_F sin\theta_C cos\theta_C |\psi(q)|^2$
is assumed.

At $X=0$ and $6\zeta=(1-\epsilon)$ these results go into
those of \cite{Zen} with $\sqrt{2}\kappa_0=b^Z$.

Note that the 2-quark PV-amplitudes of \cite{Zen06} can be written
in terms of those of \cite{Sharma} through another relation,
namely, as a superposition of the PV- and PV-amplitudes of
\cite{Sharma}, $A^{PV, Zen}=x\cdot A^{PV, Sh}+A^{PС, Sh}$; this
fact is due to the inclusion of states of different parities ($x$
being a parameter equal to zero in the exact unitary symmetry
scheme and taken equal to 1/3 in calculations with $SU(3)$
breaking \cite{Zen06}).

As already noted, putting in this formula of \cite{Zen}
$\kappa=[(1+\epsilon)c-8a]\sqrt{2}/18$,  one obtains 1-quark
transition of \cite{Sharma} (see their Table 2) at
$\sqrt{2}\kappa=b^{PV}$.

Thus, the total PV-amplitudes $A^{PV}$  are described practically
in the same way in various models as a sum of 2- and 1-quark
transition contributions. Our best fit is obtained with
$b^{PV}/3=0.96$, $d^{PV}=-0.5$ ( in units $10^{-7}\mu_N$),
$\epsilon^{PV}=-5/16$. (In the interval $|\epsilon^{PV}|\leq5/16$
the results are practically the same.)
\begin{eqnarray}
A^{PV}(\Sigma^+\rightarrow p\gamma)=
-\frac{(5+\epsilon^{PV})}{9\sqrt{2}}b^{PV}+
\frac{1}{\sqrt{2}}d^{PV}=-\frac{1}{\sqrt{2}}(1.50+0.50)=-1.41;\quad
\end{eqnarray}
\begin{eqnarray}
A^{PV}(\Sigma^0\rightarrow n\gamma)=
-\frac{(1-\epsilon^{PV})}{18}b^{PV}-
\frac{1}{2}d^{PV}=\frac{1}{2}(-0.42+0.50)=0.04;\quad
\end{eqnarray}
\begin{eqnarray}
A^{PV}(\Lambda\rightarrow n\gamma)=
\frac{(3+\epsilon^{PV})}{6\sqrt{3}}b^{PV}-
\frac{3\sqrt{3}}{2}d^{PV}=\frac{1}{\sqrt{3}}(1.35+2.25)=2.08;\quad
\end{eqnarray}
\begin{eqnarray}
A^{PV}(\Xi^0\rightarrow \Lambda\gamma)=
-\frac{(2+\epsilon^{PV})}{9\sqrt{3}}b^{PV}+
\frac{\sqrt{3}}{2}d^{PV}=\frac{1}{\sqrt{3}}(-0.54-0.75)=-0.75;\quad
\end{eqnarray}
\begin{eqnarray}
A^{PV}(\Xi^0\rightarrow \Sigma^0\gamma)=
\frac{1}{3}b^{PV}-
\frac{5}{2}d^{PV}=0.95+1.25=2.20;\quad
\end{eqnarray}
\begin{eqnarray}
A^{PV}(\Xi^-\rightarrow \Sigma^- \gamma)=
0\cdot b^{PV}+
\frac{5}{\sqrt{2}}d^{PV}=-1.75.
\end{eqnarray}
One can see that all observed decays have contributions of the
same order of magnitude to the PV-amplitudes (but the last one, as
it is obvious) from the 2- and 1-quark diagrams.

However, the case of the  PC-amplitudes proves to be more
difficult. Indeed, if in the quark model of \cite{Sharma} the
PV-amplitudes $A^{PV}$'s have the same structure as $A^{PC}$'s
(see the 3rd columns in Tables 3,4), in a series of works
\cite{Zen}-\cite{Zen06} the PC-amplitudes were analysed in terms
of pole models and unitary symmetry models, elaborated for
description of nonleptonic hyperon decays without a direct appeal
to quark models (earlier works of this kind could be found in
\cite{ZenM}):
\begin{eqnarray}
B(\Sigma^+\rightarrow p\gamma)=
\frac{\sqrt{2}}{3}(\frac{f}{d}-1)(1-\epsilon)((\frac{F}{D}-1))C
\\
B(\Sigma^0\rightarrow n\gamma)=
\frac{4}{3}C-\frac{1}{3}(\frac{f}{d}-1)(1-\epsilon)(\frac{F}{D}-1)C
\nonumber\\
B(\Lambda\rightarrow n\gamma)=
\frac{4}{3\sqrt{3}}C+
\frac{1}{9\sqrt{3}}(\frac{3f}{d}+1)(1-\epsilon)(\frac{3F}{D}+1)C
\nonumber\\
B(\Xi^0\rightarrow \Lambda\gamma)=
-\frac{4}{3\sqrt{3}}C-
\frac{1}{9\sqrt{3}}(\frac{3f}{d}-1)(1-\epsilon)(\frac{3F}{D}-1)C
\nonumber\\
B(\Xi^0\rightarrow \Sigma^0\gamma)=
-\frac{4}{3}C+\frac{1}{3}(\frac{f}{d}+1)(1-\epsilon)(\frac{F}{D}+1)C
\nonumber\\
B(\Xi^-\rightarrow \Sigma^-\gamma)=
-\frac{\sqrt{2}}{3}(\frac{f}{d}+1)(1-\epsilon)(\frac{F}{D}+1)C
\nonumber
\label{zen}
\end{eqnarray}
At first sight, it seems impossible to compare the conclusions of
this and similar models with the quark models except as in the
limit of the exact $SU(6)$ model.

Nevertheless, we shall show now in what way it is possible to
expand this expression into a sum of 1- and 2- quark contributions
for all the PC-amplitudes.

For this purpose, it is sufficient to put $f/d=-1+z$, $F/D=2/3+Z$
in the PC-amplitudes of Eq.(\ref{zen}) (see Eq.(5.2b) in
\cite{Zen}). With this step the  PC-amplitudes are splitted into
the part corresponding to 2- quark contributions (the 1st term
without $z$ and/or $Z$) and three parts corresponding to 1- quark
contributions proportional to the factors $z$, $Z$ and $z\cdot Z$.

One can easily be conviced that with $z=Z=0$ relations,
Eq.(\ref{zen}) reduces to redefinitions of the 2nd column of Table
5.

Instead, at $Z=0$ the 1-quark contributions proportional to $z$
coincide up to redefinitions with those obtained in \cite{Sharma}
and Eq.(\ref{eff}) with  $F_{ds}/D_{ds}=2/3$ (see the 2nd and the
last columns of Table 2). The terms with  $Z\not=0$ also
correspond to the 1-quark contributions but at a different
effective value of the ratio $F_{ds}/D_{ds}$ in Eq.(\ref{eff}),
namely, the terms proportional to $Z$ at $z=0$ correspond to the
choice of the effective $ds$ - current with $F_{ds}/D_{ds}=-1$,
while the terms proportional to $z\cdot Z$ correspond to pure
$F_{ds}$ - current (see the last column of Table 2).

Consider now one-by-one all the hyperon decays beginning from the
transformation of the PC-amplitude of the decay
$\Sigma^+\rightarrow p\gamma$ from \cite{Zen} and putting also our
best fit results ($C=6.3$ in units $10^{-7}\mu_N$, $z=0.05$,
$Z=0.16$, $\epsilon=-2.0$) for every contribution and their sum.
\begin{eqnarray}
B(\Sigma^+\rightarrow p\gamma)=[
\frac{4}{9\sqrt{2}}-
\frac{\sqrt{2}}{9}z-
\frac{2\sqrt{2}}{3}Z+
\frac{\sqrt{2}}{3}z\cdot Z](1-\epsilon)Z\cdot C\Rightarrow
\\
\frac{\sqrt{2}}{3}6.3 (2-0.05-0.96+0.024)=3.00.\qquad
\nonumber
\end{eqnarray}
It is seen that the main contribution comes from the 2-quark
amplitudes and from the 1-quark amplitude with the effective
$ds$-current with $F_{ds}/D_{ds}=-1$, whereas the standard 1-
quark contribution, corresponding to the choice of the effective
$ds$ - current with $F_{ds}=(2/3)D_{ds}$ is small and the last 1-
quark contribution with the pure $F_{ds}$ effective current could
be safely neglected.

Formally, we write expansion also for non-observable decay
$\Sigma^0\rightarrow n\gamma$:
\begin{eqnarray}
B(\Sigma^0\rightarrow n\gamma)=
\frac{4}{3}C-\frac{1}{3}(\frac{f}{d}-1)(1-\epsilon)(\frac{F}{D}-1)C
\nonumber\\
\Rightarrow [\frac{4(5+\epsilon)}{18}+
\frac{1}{9}z(1-\epsilon)+
\frac{2}{3}(1-\epsilon)Z-
\frac{1}{3}z(1-\epsilon)Z]\cdot C\Rightarrow
\\
\frac{1}{3} 6.3 (2.00+0.05+0.96-0.024)=6.33;\qquad
\nonumber
\end{eqnarray}
and all we have said previously about the decay
$\Sigma^+\rightarrow p\gamma$ is valid for this decay too.

Next we consider the decay $\Lambda\rightarrow n\gamma$.
\begin{eqnarray}
B(\Lambda\rightarrow n\gamma)=
\frac{4}{3\sqrt{3}}C+
\frac{1}{9\sqrt{3}}(\frac{3f}{d}+1)(1-\epsilon)(\frac{3F}{D}+1)C
\\
\Rightarrow [\frac{4(1+\epsilon)}{6\sqrt{3}}+
\frac{1}{\sqrt{3}}z(1-\epsilon)-
\frac{2}{3\sqrt{3}}(1-\epsilon)Z+
\frac{1}{\sqrt{3}}z(1-\epsilon)Z]\cdot C\Rightarrow
\nonumber\\
\frac{1}{\sqrt{3}} 6.3 (-0.667+0.15-0.32+0.024)=-2.96.\qquad \nonumber
\end{eqnarray}
It is seen that here the main contribution comes from the 2-quark
amplitudes and from the 1-quark amplitude with the effective $ds$
- current with $F_{ds}/D_{ds}=-1$, whereas the standard 1-quark
contribution is small. The last 1-quark contribution with the pure
$F_{ds}$ effective current proportional to $z\cdot Z$ from now on
is neglected.

Now let us consider radiative decays of the cascade hyperons.
\begin{eqnarray}
B(\Xi^0\rightarrow \Lambda\gamma)=
-\frac{4}{3\sqrt{3}}C-
\frac{1}{9\sqrt{3}}(\frac{3f}{d}-1)(1-\epsilon)(\frac{3F}{D}-1)C
\\
\Rightarrow [-\frac{4(2+\epsilon)}{9\sqrt{3}}-
\frac{1}{3\sqrt{3}}z(1-\epsilon)+
\frac{4}{3\sqrt{3}}(1-\epsilon)Z-
\frac{1}{\sqrt{3}}z(1-\epsilon)Z]\cdot C\Rightarrow
\nonumber\\
\frac{1}{\sqrt{3}} 6.3 (-0.05+0.64-0.024)=2.06.\qquad
\nonumber
\end{eqnarray}
Here the main contribution comes from the 1-quark amplitude with
the effective $ds$ - current with $F_{ds}/D_{ds}=-1$ while the
2-quark amplitudes vanish by the choice of $\epsilon$. This
solution is dictated by negative observed asymmetry in this decay.

The next PC-amplitude has the form:
\begin{eqnarray}
B(\Xi^0\rightarrow \Sigma^0\gamma)=
-\frac{4}{3}C+\frac{1}{3}(\frac{f}{d}+1)(1-\epsilon)(\frac{F}{D}+1)C
\\
\Rightarrow [-\frac{4}{3}+
\frac{5}{9}z(1-\epsilon)+0\cdot Z+
\frac{1}{3}z(1-\epsilon)Z]\cdot C\Rightarrow
\nonumber\\
\frac{1}{3} 6.3 (-4.0+0.25+0.024)=-7.82.\qquad
\nonumber
\end{eqnarray}
In this decay the main contribution to the PC-amplitude comes from
the 2-quark amplitudes while the 1-quark contribution with the
effective $ds$ - current with $F_{ds}/D_{ds}=-1$ is small and the
standard 1- quark  contribution corresponding to the choice of the
effective $ds$ - current with $F_{ds}=(2/3)D_{ds}$ is zero.

In conclusion we give the PC-amplitude of the decay
$\Xi^-\rightarrow \Sigma^-\gamma$.
\begin{eqnarray}
B(\Xi^-\rightarrow \Sigma^-\gamma)=
-\frac{\sqrt{2}}{3}(\frac{f}{d}+1)(1-\epsilon)(\frac{F}{D}+1)C
\\
\Rightarrow 0\cdot C -\frac{5\sqrt{2}}{9}z(1-\epsilon)C+
0\cdot Z\cdot C-\frac{\sqrt{2}}{3}z(1-\epsilon)Z\cdot C\Rightarrow
\nonumber\\
-\frac{\sqrt{2}}{3} 6.3\cdot (0.25+0.024)=-0.81, \nonumber
\end{eqnarray}
and here there are no 2-quark contributions, the 1-quark
contribution with the effective $ds$-current with
$F_{ds}/D_{ds}=-1$ is zero while the standard 1-quark term gives
the main contribution.

So the 1st term in every formula for the PC-amplitudes (i.e., for
PC-amplitudes at $z=Z=0$ or, which is the same, at $f/d=-1$,
$F/D=2/3$) corresponds exactly to the expressions in \cite{Sharma}
at $X=0$ (see Table 4); this fact was partially noted in
\cite{Zen}. The 1-quark contributions, proportional to $z$, also
coincide with those in \cite{Sharma}) and  \cite{Zen} (one should
put $\sqrt{2}/3 z(1-\epsilon)C\equiv a$, $a$ from Table 2).
However, and it is important, if in the quark model of
\cite{Sharma} the PV- and PC- amplitudes had identical factors of
the 2-quark terms, in \cite{Zen} and the present one this is not
the case. (Even more, the parameter $\epsilon$ is negative here
and different for the PC- and PC-amplitudes.) It can be thought
that in this way phenomenological models effectively take into
account the difference between the dynamics of the processes going
with and without parity conservations. Our results are given in
Table 6.

Unfortunately, at the present time, it seems to be impossible to
perform dynamical calculations in an unambiguous way.

\section{Conclusion}

It is shown that many models describing weak radiative hyperon
decays can be reduced to rather a simple quark model including
1-quark transitions with the effective $sd\gamma$- vertex and
2-quark process with the $W$ -- exchange $s+u\rightarrow
u+d+\gamma$. As an example, quark and unitary models \cite{Sharma}
and \cite{Zen} are considered. Using them as a basis, a
phenomenological model is constructed which describes the data and
gives clear predictions for the asymmetry parameters of the decays
$\Lambda\rightarrow n+\gamma$ and $\Xi^-\rightarrow \Sigma^-
+\gamma$.

For the 2-quark processes with the $W$ -- exchange rather general
expressions are obtained which could be used not only for the
hyperon decays but also for the decay of the new heavy baryons.

We do not discuss here a traditional problem connected with the
Hara theorem prediction of zero asymmetry in the decay $\Sigma^+
\rightarrow p+\gamma$, as already in the GIM model this problem
can be overcome \cite{BukGIM}.

Nevertheless, there are many theoretical problems unresolved in
the models of weak radiative hyperon decays (together with those
of hyperon nonleptonic decays) which expect a more thorough
analysis.

\newpage

{\bf Appendix A }

{The 2-quark contributions to neutral hyperon decays
}
\vskip 5mm
 \begin{enumerate}
\item
Let us analyse the decay $ \Sigma^{0} \rightarrow n \gamma $,
although it cannot be seen soon experimentally. The  2-quark
amplitude of this decay can be expressed via the following matrix
elements:
$$
6\sqrt{2}<n_{\downarrow}, \gamma(+1) |O| \Sigma^{0}_{\uparrow}>=
\qquad \qquad (А.1)
$$
$$=<2d_{2}d_{2}u_{1}-d_{2}u_{2}d_{1}-
u_{2}d_{2}d_{1},\gamma(+1)|O|2u_{1}d_{1}s_{2}+2d_{1}u_{1}s_{2}-$$
$$-u_{1}s_{1}d_{2}-s_{1}u_{1}d_{2}-d_{1}s_{1}u_{2}-s_{1}d_{1}u_{2}>=$$
$$=8<d_{2}d_{2}u_{1},\gamma(+1) |O|u_{1}d_{1}s_{2}>-
8<d_{2}d_{2}u_{1},\gamma(+1) |O|u_{1}s_{1}d_{2}>- $$
$$-4<d_{2}d_{2}u_{1},\gamma(+1) |O|d_{1}s_{1}u_{2}>-
8<d_{2}u_{2}d_{1},\gamma(+1) |O|u_{1}d_{1}s_{2}>+$$
$$+4<d_{2}u_{2}d_{1},\gamma(+1) |O|u_{1}s_{1}d_{2}>+
4<d_{2}u_{2}d_{1},\gamma(+1) |O|d_{1}s_{1}u_{2}>$$
The 1st and  2nd matrix elements (m.e.'s ) in the RHS of this
expression correspond to the 2nd and 3rd diagrams of Fig.1, so we
do not consider them (see \cite{Verma}).
The 2nd m.e. in the RHS of Eq.(A.1)
$ <d_{2}d_{2}u_{1}, \gamma(+1) |O|u_{1}s_{1}d_{2}>= A_{2} $ is
described by three diagrams of Fig.4 but with the quark $ d_{2}$
as a spectator. To the 4th m.e. in the RHS of Eq.(A.1) $
<d_{2}u_{2}d_{1}, \gamma(+1) |O|u_{1}d_{1}s_{2}>= A_{3} $ three
diagrams of Fig.3 contribute but with the quark $ d_{1}$ as a
spectator.

To the 5th m.e. in the RHS of Eq.(A.1) $ <d_{2}u_{2}d_{1},
\gamma(+1) |O|u_{1}s_{1}d_{2}>= A_{1} $ three diagrams of Fig.2
contribute but with the quark $ d_{2}$ as a spectator.

To the 6th m.e. in the RHS of  Eq.(A.1) $
<d_{2}u_{2}d_{1},\gamma(+1) |O|d_{1}s_{1}u_{2}>=A_{4} $ three
diagrams of Fig.5 contribute but with the quark $ d_{1}$ as a
spectator. Their sum gives
$$ <n_{\downarrow}, \gamma(+1) |O| \Sigma^{0}_{\uparrow} >=
\frac{2}{3\sqrt{2}}(A_{1}-2A_{2}-2A_{3}+A_{4}).$$
\item
Now let us describe the decay $ \Lambda \rightarrow n \gamma $:
$$
 2 \sqrt{6}<n_{\downarrow}, \gamma(+1) |O| \Lambda_{\uparrow} >=
\qquad \qquad (А.2)
$$
$$= <2d_{2}d_{2}u_{1}-d_{2}u_{2}d_{1}-u_{2}d_{2}d_{1},\gamma(+1)|O|
u_{1}s_{1}d_{2}+s_{1}u_{1}d_{2}-d_{1}s_{1}u_{2}-s_{1}d_{1}u_{2}>=$$
$$= 4<d_{2}d_{2}u_{1},\gamma(+1)|O|u_{1}s_{1}d_{2}>-
4<d_{2}d_{2}u_{1},\gamma(+1)|O|d_{1}s_{1}u_{2}>-$$
$$- 4<d_{2}u_{2}d_{1},\gamma(+1)|O|u_{1}s_{1}d_{2}>+
4<u_{2}d_{2}d_{1},\gamma(+1)|O|d_{1}s_{1}u_{2}>.$$
The 1st m.e. in the RHS of Eq.(A.2) $ <d_{2}d_{2}u_{1},\gamma(+1)
|O|u_{1}s_{1}d_{2}>=A_{2} $ is given by the contributions of three
diagrams of Fig.4 with the spectator $ d_{2}$. The 2nd m.e. in the
RHS of Eq.(A.2) $ <d_{2}d_{2}u_{1},\gamma(+1) |O|d_{1}s_{1}u_{2}>$
in the case of $W$-exchange between quarks can be described by the
diagram of Fig.1 and should be very small. To the 3rd m.e. in the
RHS of Eq.(A.2) $ <d_{2}u_{2}d_{1},\gamma(+1)
|O|u_{1}s_{1}d_{2}>=A_{1} $ three diagrams of Fig.2 contribute but
with the quark $ d_{2}$ as a spectator.

To the 4th m.e. in the RHS of  Eq.(A.2) $
<d_{2}u_{2}d_{1},\gamma(+1) |O|d_{1}s_{1}u_{2}>=A_{4} $ three
diagrams of Fig.5 contribute but with the quark $ d_{1}$ as a
spectator.

Finally one obtains:
$$ <n_{\downarrow}, \gamma(+1) |O| \Lambda_{\uparrow} >=
\frac{2}{\sqrt{6}}(A_{1}-2A_{2}-A_{4}).$$
\item
For the decay $ \Xi^{0} \rightarrow \Lambda \gamma $ we have
$$
 2 \sqrt{6}<\Lambda_{\downarrow}, \gamma(+1) |O| \Xi^{0}_{\uparrow} >=
\qquad \qquad (А.3)
$$
$$= <u_{2}s_{2}d_{1}+s_{2}u_{2}d_{1}-d_{2}s_{2}u_{1}-s_{2}d_{2}u_{1},
\gamma(+1)|O|
2s_{1}s_{1}u_{2}-s_{1}u_{1}s_{2}-u_{1}s_{1}s_{2}>=$$
$$= 4<u_{2}s_{2}d_{1}, \gamma(+1) |O|s_{1}s_{1}u_{2}>-
4<u_{2}s_{2}d_{1}, \gamma(+1) |O|s_{1}u_{1}s_{2}>-$$
$$-4<d_{2}s_{2}u_{1}, \gamma(+1) |O|s_{1}s_{1}u_{2}>+
4<d_{2}s_{2}u_{1}, \gamma(+1) |O|s_{1}u_{1}s_{2}>.$$
The 1st and the 3rd matrix elements (m.e.'s ) in the RHS of this
expression correspond to the 2nd and 3rd diagrams of  Fig.1, and
we neglect them both. The 2nd m.e. in the RHS of Eq.(A.3) $
<u_{2}s_{2}d_{1}, \gamma(+1) |O|s_{1}u_{1}s_{2}>= A_{1}$ is given
by the contributions of three diagrams of Fig.2 with the spectator
$ s_{2}$. To the 4th m.e. in the RHS of  Eq.(A.3)
$<d{2}s_{2}u_{1}, \gamma(+1) |O|s_{1}u_{1}s_{2}>= A_{2}$ three
diagrams of Fig.4 contribute but with the quark $ s_{2}$ as a
spectator. Finally,
$$ <\Lambda_{\downarrow}, \gamma(+1) |O| \Xi^{0}_{\uparrow} >=
\frac{2}{\sqrt{6}}(A_{1}-A_{2}).$$
\item
For the decay $ \Xi^{0} \rightarrow \Sigma^{0} \gamma $
one has
$$
6 \sqrt{2}< \Sigma^{0}_{ \downarrow}, \gamma(+1) |O| \Xi^{0}_{\uparrow}>=
<2u_{2}d_{2}s_{1}+2d_{2}u_{2}s_{1} -
\qquad \qquad (А.4)
$$
 $$-u_{2}s_{2}d_{1}-s_{2}u_{2}d_{1}-d_{2}s_{2}u_{1}-s_{2}d_{2}u_{1},
\gamma(+1)|O|
2s_{1}s_{1}u_{2}-s_{1}u_{1}s_{2}-u_{1}s_{1}s_{2}>=$$
$$8<u_{2}d_{2}s_{1}, \gamma(+1) |O|s_{1}s_{1}u_{2}>-
8<u_{2}d_{2}s_{1}, \gamma(+1) |O|s_{1}u_{1}s_{2}>-$$
$$-4<u_{2}s_{2}d_{1}, \gamma(+1) |O|s_{1}s_{1}u_{2}>+
4<u_{2}s_{2}d_{1}, \gamma(+1) |O|s_{1}u_{1}s_{2}>- $$
$$- 4<d_{2}s_{2}u_{1}, \gamma(+1) |O|s_{1}s_{1}u_{2}>+
4<d_{2}s_{2}u_{1}, \gamma(+1) |O|s_{1}u_{1}s_{2}>.$$
The 1st m.e. in the RHS of  Eq.(A.4) $ <u_{2}d_{2}s_{1},
\gamma(+1) |O|s_{1}s_{1}u_{2}>=A_{4} $ is described by three
diagrams of Fig.5 with the spectator $ s_{1}$.

The 2nd m.e. in the RHS of Eq.(A.4) $ <u_{2}d_{2}s_{1}, \gamma(+1)
|O|s_{1}u_{1}s_{2}>=A_{3}$ is given by the contributions of three
diagrams of Fig.3 with the spectator $ s_{1}$.

The 3rd and the 5th matrix elements (m.e.'s ) in the RHS of this
expression correspond to the diagrams of Fig.1 type, and we
neglect them both.

To the 4th m.e. in the RHS of  Eq.(A.4) $ <u_{2}s_{2}d_{1},
\gamma(+1) |O|s_{1}u_{1}s_{2}>=A_{1}$ three diagrams of Fig.2
contribute but with the quark $ s_{2}$ as a spectator. And,
finally, to the 6th m.e. in the RHS of Eq.(A.4) $
<d_{2}s_{2}u_{1}, \gamma(+1) |O|s_{1}u_{1}s_{2}>=A_{2}$ three
diagrams of Fig.4 contribute but with the quark $ s_{2}$ as a
spectator. Their sum gives:
$$ <\Sigma^{0}_{\downarrow}, \gamma(+1) |O| \Xi^{0}_{\uparrow} >=
\frac{2}{3 \sqrt{2}}(A_{1}+A_{2}-2A_{3}+4A_{4}) .$$
\end{enumerate}
The analysis performed shows that decays of all the neutral
hyperons as well as that of $ \Sigma^{+} \rightarrow p \gamma $
can be expressed in terms of the same amplitudes $ A_{k},\, k=1,
2, 3, 4,$ assumig that spectator quarks do not change them.

\newpage
{\bf Appendix B}

{Relation between two different representations of the
PC-amplitudes} \vskip 5mm The form of the PC-amplitudes considered
here is not obviously unique in the framework of unitary symmetry
models. In one of the recent works \cite{Zen06} another form was
used which was close to the old form presented by \cite{Pak}:
$$
B(\Sigma^+\rightarrow p\gamma)=
\sqrt{2}(\frac{f}{d}-1)(\mu_{\Sigma^+}-\mu_p)\frac{N}{\mu_p},
\qquad \qquad (Б.1)
$$
$$
B(\Sigma^0\rightarrow n\gamma)=
[-(\frac{f}{d}-1)(\mu_{\Sigma^0}-\mu_n)+
\frac{1}{\sqrt{3}}(\frac{3f}{d}+1)\mu_{\Sigma\Lambda}]
\frac{N}{\mu_p},
$$
$$
B(\Lambda\rightarrow n\gamma)=
[\frac{1}{\sqrt{3}}(\frac{3f}{d}+1)(\mu_{\Lambda}-\mu_n)-
(\frac{f}{d}-1)\mu_{\Sigma\Lambda}]
\frac{N}{\mu_p},
$$
$$
B(\Xi^0\rightarrow \Lambda\gamma)=
[-\frac{1}{\sqrt{3}}(\frac{3f}{d}-1)(\mu_{\Xi^0}-\mu_{\Lambda})-
(\frac{f}{d}+1)\mu_{\Sigma\Lambda}]
\frac{N}{\mu_p},
$$
$$
B(\Xi^0\rightarrow\Sigma^0\gamma)=
[(\frac{f}{d}+1)(\mu_{\Sigma^0}-\mu_{\Sigma^0})+
\frac{1}{\sqrt{3}}(\frac{3f}{d}-1)\mu_{\Sigma\Lambda}]
\frac{N}{\mu_p},
$$
$$
B(\Xi^-\rightarrow\Sigma^-\gamma)=
-\sqrt{2}(\frac{f}{d}+1)(\mu_{\Xi^-}-\mu_{\Sigma^-})
\frac{N}{\mu_p}.
$$
They can be related between them, as was noted in \cite{Zen}, with
the help of the simple representation of the broken unitary model
of the baryon magnetic moments (We give also values of the
magnetic moments which could be obtained in this simple model at
$F=1.8$, $D=2.7$ и $(1-\epsilon)=1/3$):
$$
\mu_p=F+\frac{1}{3}D=2.7,
\qquad \qquad (Б.2)
$$
$$
\mu_n=-\frac{2}{3}D=-1.8,
$$
$$
\mu_{\Sigma^+}=F+\frac{1}{3}D+\frac{1}{3}(1-\epsilon)(F-D)=2.6,
$$
$$
\mu_{\Sigma^-}=-F+\frac{1}{3}D+\frac{1}{3}(1-\epsilon)(F-D)=-1.0,
$$
$$
\mu_{\Xi^-}=-F+\frac{1}{3}D+\frac{2}{3}(1-\epsilon)F-0.5,
$$
$$
\mu_{\Xi^0}=-\frac{2}{3}D+\frac{2}{3}(1-\epsilon)F=-1.4,
$$
$$
\mu_{\Lambda}=-\frac{1}{3}D+\frac{1}{9}(1-\epsilon)(3F+D)=-0.6.
$$
However, this representation does not lead to a better description
of the data and does not change our conclusion as to strong
differences in the coefficients in the description of the PV- and
PC- amplitudes.

\newpage
\pagestyle{empty}

Table 1. Weak radiative hyperon decays (WRHD),
experiment~\cite{PDG}, BR is the branching ratio of the radiative
decay, $ \Gamma_\gamma$ is the radiative partial width, and
$A_\gamma$ is the asymmetry parameter of teh WRHD. \vspace{5mm}
\begin{figure}
\begin{center}

\begin{tabular}{|c|c|c|c|c|} \hline
& & & & \\
Decay                &  BR $(\times 10^{3}) $
& $ \Gamma_\gamma\times 10^{+15}$ МэВ
&    $ A_\gamma$   & $k_\gamma^3\times 10^3  $ ГэВ$^3$       \\
& & & & \\ \hline
& & & & \\
$\Sigma^{+} \rightarrow p \gamma$        & $1,23 \pm 0,05$
& $10.25\pm 0.40 $ &  $-0,76 \pm 0,08$ & 11.4  \\
& & & &  \\ \hline
& & & & \\
$\Sigma^{0} \rightarrow n \gamma$        & $-$
&- &       $-$       & 11.6 \\
& & & &     \\ \hline
& & & & \\
$\Lambda^{0} \rightarrow n \gamma$       & $1,75 \pm 0,15$
& $4.43\pm 0.40$ &       $-$ &4.25     \\
& & & &       \\ \hline
& & & & \\
$\Xi^{0} \rightarrow \Lambda \gamma$     & $1,16 \pm 0,08$
& $2.67\pm 0.20$ &  $-0,78 \pm 0,19 $ &6.23   \\
& & & & \\ \hline
& & & & \\
$\Xi^{0} \rightarrow \Sigma^{0} \gamma$  & $3,33 \pm 0,10$
& $7.65\pm 0.19$ &  $-0,63 \pm 0,09 $  &1.60  \\
& & & & \\ \hline
& & & & \\
$\Xi^{-} \rightarrow \Sigma^{-} \gamma$  & $0,127 \pm 0,023$
& $0.502 \pm 0.090 $ &  $-$  & 1.64     \\
& & & & \\ \hline
\end{tabular}
\end{center}
\end{figure}
\vskip 5mm

\newpage

Table 2. Contributions of the 1-quark diagrams and of the
effective strangeness-changing neutral  $SU(3)_{f}$ current to
WRHD. \vspace{5mm}
\begin{figure}[t]
\begin{center}
\begin{tabular}{|c|c|c|c|} \hline
&  & & \\
Decay  &  \cite{Sharma} &
Eq.(\ref{new})
& Eq.(\ref{eff})\\
& & & \\ \hline
&  & & \\
$ \Sigma^{+}\rightarrow p \gamma$  & $-b/3 $ & $0$
& $-F_{ds}+D_{ds}$ \\
&  & & \\ \hline
&  & & \\
$ \Sigma^{0}\rightarrow n \gamma$  & $b/3\sqrt{2}  $
& $a^{d}/\sqrt{2} $
& $(F_{ds}-D_{ds})/\sqrt{2}$  \\
& & & \\ \hline
& & & \\
$\Lambda^{0}\rightarrow n \gamma$  & $3b/\sqrt{6}$
& $a^{d}/\sqrt{6}  $
& $-(3F_{ds}+D_{ds})/\sqrt{6}$ \\
 & & & \\ \hline
 & & & \\
$\Xi^{0}\rightarrow \Lambda \gamma$  & $b/\sqrt{6}$
& $-a^{d}/\sqrt{6}$
& $(3F_{ds}-D_{ds})/\sqrt{6}$  \\
 & & & \\ \hline
 & & & \\
$\Xi^{0}\rightarrow \Sigma^{0} \gamma$  & $ -5b/3\sqrt{2}$
& $ -a^{d}/\sqrt{2} $
& $ -(F_{ds}+D_{ds})/\sqrt{2}$ \\
 & & & \\ \hline
 & & & \\
$ \Xi^{-}\rightarrow \Sigma^{-} \gamma$  & $5b/3$ & $0$
& $F_{ds}+D_{ds}$\\
 & & & \\ \hline
\end{tabular}
\end{center}
\end{figure}

\newpage

Table 3. WRHD, 2-quark diagram contributions to PV-amplitudes
\vspace{5mm}
\begin{figure}
\begin{center}
\begin{tabular}{|c|c|c|c|} \hline
 & & & \\
Decay &  \cite{Zen} & \cite{Sharma}
&  Eq.(\ref{AAA})  \\
 & & & \\ \hline
 & & & \\
$ \Sigma^{+}\rightarrow p \gamma$  & $-\frac{5+\epsilon}{9\sqrt{2}}b$
& $ \frac{2}{9}[-3-2X+\zeta(3+X)]$
& $ \frac{2}{3}(-2A_{1}^{PV}+A_{2}^{PV}-2A_{3}^{PV}+A_{4}^{PV})$ \\
 & & & \\ \hline
 & & & \\
$ \Sigma^{0}\rightarrow n \gamma$  & $-\frac{1-\epsilon}{18}b  $
& $ \frac{2}{9\sqrt{2}}[-2X+\zeta(-3+X)] $
& $ \frac{2}{3\sqrt{2}}(A_{1}^{PV}-2A_{2}^{PV}-2A_{3}^{PV}+A_{4}^{PV})$\\
 & & & \\ \hline
 & & & \\
$ \Lambda^{0}\rightarrow n \gamma$  & $\frac{3+\epsilon}{6\sqrt{3}}b$
& $ \frac{2}{3\sqrt{6}}[-2+\zeta(-3+X)] $
& $ \frac{2}{\sqrt{6}}(A_{1}^{PV}-2A_{2}^{PV}-A_{4}^{PV}) $ \\
 & & & \\ \hline
 & & & \\
$\Xi^{0}\rightarrow \Lambda \gamma$  & $-\frac{2+\epsilon}{9\sqrt{3}}b$
& $ \frac{2}{3\sqrt{6}}[1-2\zeta] $
& $ \frac{2}{\sqrt{6}}(A_{1}^{PV}-A_{2}^{PV})$  \\
 & & & \\ \hline
 & & & \\
$ \Xi^{0}\rightarrow \Sigma^{0} \gamma $  & $ \frac{1}{3}b $
& $ \frac{2}{9 \sqrt{2}}[-3-2X-2\zeta X] $
& $ \frac{2}{3 \sqrt{2}}(A_{1}^{PV}+A_{2}^{PV}-2A_{3}^{PV}+4A_{4}^{PV})$ \\
 & & & \\ \hline
 & & & \\
$ \Xi^{-}\rightarrow \Sigma^{-} \gamma$  & $0$ & $0$ & $0$ \\
 & & & \\ \hline
\end{tabular}
\end{center}
\end{figure}
\newpage

Table 4. WRHD, 2-quark diagram contributions to PC-amplitudes

\vspace{5mm}
\begin{figure}
\begin{center}
\begin{tabular}{|c|c|c|c|} \hline
 & & & \\
Decay &  \cite{Zen} & \cite{Sharma}
& Eq.(\ref{AAA})  \\
 & & & \\ \hline
 & & & \\
$ \Sigma^{+}\rightarrow p \gamma$  & $-\frac{1-\epsilon}{9\sqrt{2}}b$
& $ \frac{2}{9}[X+\zeta(3+X)]$
& $ \frac{2}{3}(-2A_{1}^{PC}+A_{2}^{PC}-2A_{3}^{PC}+A_{4}^{PC})$ \\
 & & & \\ \hline
 & & & \\
$ \Sigma^{0}\rightarrow n \gamma$  & $-\frac{5+\epsilon}{18}b  $
& $ \frac{2}{9\sqrt{2}}[3+X+\zeta(-3+X)] $
& $ \frac{2}{3\sqrt{2}}
(A_{1}^{PC}-2A_{2}^{PC}-2A_{3}^{PC}+A_{4}^{PC})$\\
 & & & \\ \hline
 & & & \\
$ \Lambda^{0}\rightarrow n \gamma$  & $\frac{1+\epsilon}{6\sqrt{3}}b$
& $ \frac{2}{3\sqrt{6}}[1-X+\zeta(-3+X)] $
& $ \frac{2}{\sqrt{6}}
(A_{1}^{PC}-2A_{2}^{PC}-A_{4}^{PC}) $ \\
 & & & \\ \hline
 & & & \\
$\Xi^{0}\rightarrow \Lambda \gamma$  & $\frac{2+\epsilon}{9\sqrt{3}}b$
& $ \frac{2}{3\sqrt{6}}[1-2\zeta] $
& $ \frac{2}{\sqrt{6}}(A_{1}^{PC}-A_{2}^{PC})$  \\
 & & & \\ \hline
 & & & \\
$ \Xi^{0}\rightarrow \Sigma^{0} \gamma $  & $ \frac{1}{3}b $
& $ \frac{2}{9 \sqrt{2}}[3+4X-2\zeta X] $
& $ \frac{2}{3 \sqrt{2}}
(A_{1}^{PC}+A_{2}^{PC}-2A_{3}^{PC}+4A_{4}^{PC})$ \\
 & & & \\ \hline
 & & & \\
$ \Xi^{-}\rightarrow \Sigma^{-} \gamma$  & $0$ & $0$ & $0$ \\
 & & & \\ \hline
\end{tabular}
\end{center}
\end{figure}

\newpage

Table 5. WRHD, phenomenological model and experiment \cite{PDG}.
Amplitudes $A^{PV}$ and $B^{PC}$ are in units of $10^{-7}\mu_N$,
$\pi \Gamma_\gamma/k_\gamma^3=|A^{PV}|^2+|B^{PC}|^2$ in units of
$(10^{-7}\mu_N)^2$. \vspace{5mm}
\begin{figure}
\begin{center}
\begin{tabular}{|c|c|c|c|c|c|} \hline
& & & & & \\
Deacy   & $A^{PV}$ & $B^{PC}$
& $\pi \Gamma_\gamma/k_\gamma^3$
&   $A_\gamma$  & $k_\gamma^3\times 10^3  $ ГэВ$^3$       \\
& & & & & \\ \hline
& & & & & \\
$\Sigma^{+} \rightarrow p \gamma$  & -1.41 & 3.00
& $11.4 $ & -0.74  & 11.4  \\
& & &($11.0\pm 0.4)^{exp.}$ & $( -0,76 \pm 0,08)^{э}$ &  \\ \hline
& & & & & \\
$\Sigma^{0} \rightarrow n \gamma$  & 0.04  & 6.27
&   40.0 & 0.01   & 11.6 \\
& & & & &     \\ \hline
& & & & & \\
$\Lambda^{0} \rightarrow n \gamma$   & 2.08   & -2.95
& 13.0 & -0.94 & 4.25     \\
& & &$(13.0\pm 1.1)^{exp.}$ & &       \\ \hline
& & & & & \\
$\Xi^{0} \rightarrow \Lambda \gamma$  & -0.75  &  2.06
& 4.08 & -0.64 & 6.23   \\
& & &$(5.4\pm 0.4)^{exp.}$ & $(-0,78 \pm 0,19)^{э} $ & \\ \hline
& & & & & \\
$\Xi^{0} \rightarrow \Sigma^{0} \gamma$& 2.20 & -7.82
& 66.0 &  -0.52  &1.60  \\
& & & $(59.75\pm 2.0)^{exp.}$ & $(-0,63 \pm 0,09)^{э} $ & \\ \hline
& & & & & \\
$\Xi^{-} \rightarrow \Sigma^{-} \gamma$ & -1.75 & -0.81
& 3.7 &  0.76  & 1.64     \\
& & & $(3.82\pm 0.8)^{exp.}$ & & \\ \hline
\end{tabular}
\end{center}
\end{figure}

\newpage

\begin{center}
\begin{picture}(400,100)(0,0)
\Line(0,60)(100,60)
\Line(0,40)(100,40)
\Line(0,20)(100,20)
\Photon(50,40)(50,60)3 3
\Photon(50,20)(50,0)3 3
\Text(5,70)[c]{$ s_{2} $}
\Text(5,50)[c]{$ u_{1} $}
\Text(5,30)[c]{$ u_{1} $}
\Text(95,70)[c]{$ u_{2} $}
\Text(95,50)[c]{$ d_{1} $}
\Text(95,30)[c]{$ u_{2} $}
\Text(40,50)[c]{$ W $}
\Text(60,0)[c]{$ \gamma $}
\Line(120,60)(220,60)
\Line(120,40)(220,40)
\Line(120,20)(220,20)
\Photon(170,40)(170,60)3 3
\Photon(170,0)(170,20)3 3
\Text(125,70)[c]{$ u_{1} $}
\Text(125,50)[c]{$ s_{2} $}
\Text(125,30)[c]{$ d_{1} $}
\Text(215,70)[c]{$ d_{2} $}
\Text(215,50)[c]{$ u_{1} $}
\Text(215,30)[c]{$ d_{2} $}
\Text(160,50)[c]{$ W $}
\Text(180,0)[c]{$ \gamma $}
\Line(240,60)(340,60)
\Line(240,40)(340,40)
\Line(240,20)(340,20)
\Photon(290,40)(290,60)3 3
\Photon(290,20)(290,0)3 3
\Text(245,70)[c]{$ s_{1} $}
\Text(245,50)[c]{$ u_{2} $}
\Text(245,30)[c]{$ d_{1} $}
\Text(335,70)[c]{$ u_{1} $}
\Text(335,50)[c]{$ d_{2} $}
\Text(335,30)[c]{$ d_{2} $}
\Text(280,50)[c]{$ W $}
\Text(300,0)[c]{$ \gamma $}
\end{picture}
\end{center}
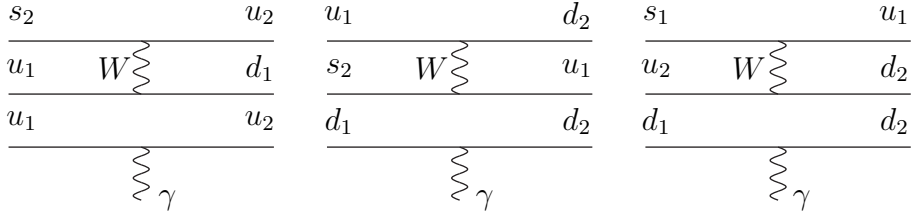
Fig. 1.  The 3-quark diagrams without
spectator quark
($q_{1}$ means $q_{\uparrow}$, $q_{2}$ means
$q_{\downarrow}$, $q=u,d,s$)

\vskip 15mm



\begin{center}
\begin{picture}(400,100)(0,0)
\Line(0,60)(100,60)
\Line(0,40)(100,40)
\Line(0,10)(100,10)
\Photon(40,40)(40,60)3 3
\Photon(70,60)(70,80)3 3
\Text(5,70)[c]{$ s_{1} $}
\Text(5,50)[c]{$ u_{1} $}
\Text(5,20)[c]{$ u_{2} $}
\Text(95,70)[c]{$ u_{2} $}
\Text(95,50)[c]{$ d_{1} $}
\Text(50,70)[c]{$ u_{1} $}
\Text(95,20)[c]{$ u_{2} $}
\Text(30,50)[c]{$ W $}
\Text(80,80)[c]{$ \gamma $}
\Line(120,60)(220,60)
\Line(120,40)(220,40)
\Line(120,10)(220,10)
\Photon(180,40)(180,60)3 3
\Photon(160,60)(160,80)3 3
\Text(125,70)[c]{$ s_{1} $}
\Text(170,70)[c]{$ s_{2} $}
\Text(125,50)[c]{$ u_{1} $}
\Text(125,20)[c]{$ u_{2} $}
\Text(215,70)[c]{$ u_{2} $}
\Text(215,50)[c]{$ d_{1} $}
\Text(215,20)[c]{$ u_{2} $}
\Text(170,50)[c]{$ W $}
\Text(150,80)[c]{$ \gamma $}
\Line(240,60)(340,60)
\Line(240,40)(340,40)
\Line(240,10)(340,10)
\Photon(300,40)(300,60)3 3
\Photon(270,20)(270,40)3 3
\Text(245,70)[c]{$ s_{1} $}
\Text(245,50)[c]{$ u_{1} $}
\Text(290,50)[c]{$ u_{2} $}
\Text(245,20)[c]{$ u_{2} $}
\Text(335,70)[c]{$ u_{2} $}
\Text(335,50)[c]{$ d_{1} $}
\Text(335,20)[c]{$ u_{2} $}
\Text(315,50)[c]{$ W $}
\Text(280,30)[c]{$ \gamma $}
\end{picture}
\end{center}

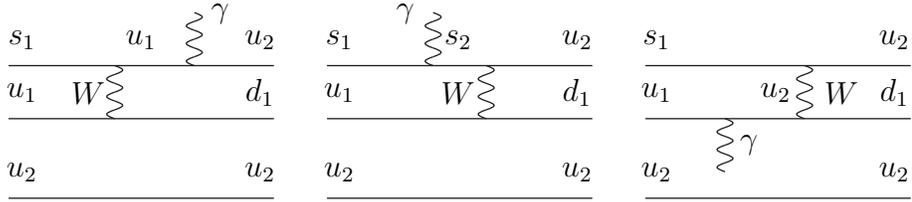
Fig. 2.  The 3-quark diagrams of the decay $ \Sigma^{+}\rightarrow
p \gamma$ corresponding to the matrix element $A_1$ with the 3rd
quark $u_{2}=u_{\downarrow}$ as a spectator.

\vskip 15mm


\begin{center}
\begin{picture}(400,100)(0,0)
\Line(0,60)(100,60)
\Line(0,40)(100,40)
\Line(0,10)(100,10)
\Photon(40,40)(40,60)3 3
\Photon(70,20)(70,40)3 3
\Text(5,70)[c]{$ s_{2} $}
\Text(5,50)[c]{$ u_{1} $}
\Text(5,20)[c]{$ u_{1} $}
\Text(95,70)[c]{$ u_{2} $}
\Text(95,50)[c]{$ d_{2} $}
\Text(50,50)[c]{$ d_{1} $}
\Text(95,20)[c]{$ u_{1} $}
\Text(30,50)[c]{$ W $}
\Text(80,30)[c]{$ \gamma $}
\Line(120,60)(220,60)
\Line(120,40)(220,40)
\Line(120,10)(220,10)
\Photon(160,40)(160,60)3 3
\Photon(190,60)(190,80)3 3
\Text(125,70)[c]{$ s_{2} $}
\Text(170,70)[c]{$ u_{1} $}
\Text(125,50)[c]{$ u_{1} $}
\Text(125,20)[c]{$ u_{1} $}
\Text(215,70)[c]{$ u_{2} $}
\Text(215,50)[c]{$ d_{2} $}
\Text(215,20)[c]{$ u_{1} $}
\Text(175,50)[c]{$ W $}
\Text(200,80)[c]{$ \gamma $}
\Line(240,60)(340,60)
\Line(240,40)(340,40)
\Line(240,10)(340,10)
\Photon(300,40)(300,60)3 3
\Photon(260,20)(260,40)3 3
\Text(245,70)[c]{$ s_{2} $}
\Text(245,50)[c]{$ u_{1} $}
\Text(280,50)[c]{$ u_{2} $}
\Text(245,20)[c]{$ u_{1} $}
\Text(335,70)[c]{$ u_{2} $}
\Text(335,50)[c]{$ d_{2} $}
\Text(335,20)[c]{$ u_{1} $}
\Text(315,50)[c]{$ W $}
\Text(270,30)[c]{$ \gamma $}
\end{picture}
\end{center}

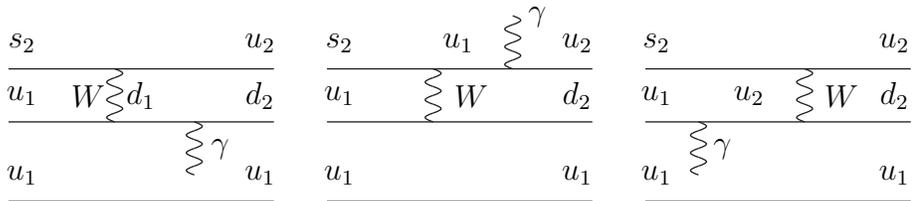
Fig. 3.  The 3-quark diagrams of the decay $ \Sigma^{+}\rightarrow
p \gamma$ corresponding to the matrix element $A_3$ with the 3rd
quark $u_{1}=u_{\uparrow}$ as a spectator.


\newpage


\begin{center}
\begin{picture}(400,100)(0,0)
\Line(0,60)(100,60)
\Line(0,40)(100,40)
\Line(0,10)(100,10)
\Photon(40,40)(40,60)3 3
\Photon(80,20)(80,40)3 3
\Text(5,70)[c]{$ s_{1} $}
\Text(5,50)[c]{$ u_{1} $}
\Text(5,20)[c]{$ u_{2} $}
\Text(95,70)[c]{$ u_{1} $}
\Text(95,50)[c]{$ d_{2} $}
\Text(60,50)[c]{$ d_{1} $}
\Text(95,20)[c]{$ u_{2} $}
\Text(30,50)[c]{$ W $}
\Text(70,20)[c]{$ \gamma $}
\Line(120,60)(220,60)
\Line(120,40)(220,40)
\Line(120,10)(220,10)
\Photon(180,40)(180,60)3 3
\Photon(160,60)(160,80)3 3
\Text(125,70)[c]{$ s_{1} $}
\Text(170,70)[c]{$ s_{2} $}
\Text(125,50)[c]{$ u_{1} $}
\Text(125,20)[c]{$ u_{2} $}
\Text(215,70)[c]{$ u_{1} $}
\Text(215,50)[c]{$ d_{2} $}
\Text(215,20)[c]{$ u_{2} $}
\Text(170,50)[c]{$ W $}
\Text(150,80)[c]{$ \gamma $}
\Line(240,60)(340,60)
\Line(240,40)(340,40)
\Line(240,10)(340,10)
\Photon(300,40)(300,60)3 3
\Photon(270,20)(270,40)3 3
\Text(245,70)[c]{$ s_{1} $}
\Text(245,50)[c]{$ u_{1} $}
\Text(290,50)[c]{$ u_{2} $}
\Text(245,20)[c]{$ u_{2} $}
\Text(335,70)[c]{$ u_{1} $}
\Text(335,50)[c]{$ d_{2} $}
\Text(335,20)[c]{$ u_{2} $}
\Text(315,50)[c]{$ W $}
\Text(280,30)[c]{$ \gamma $}
\end{picture}
\end{center}

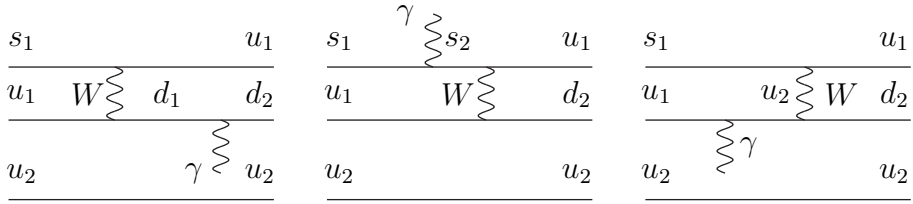
Fig. 4.  The 3-quark diagrams of the decay $ \Sigma^{+}\rightarrow
p \gamma$ corresponding to the matrix element $A_2$ with the 3rd
quark $u_{2}=u_{\downarrow}$ as a spectator.


\vskip 15mm



\begin{center}
\begin{picture}(400,100)(0,0)
\Line(0,60)(100,60)
\Line(0,40)(100,40)
\Line(0,10)(100,10)
\Photon(40,60)(40,80)3 3
\Photon(65,60)(65,40)3 3
\Text(5,70)[c]{$ s_{1} $}
\Text(5,50)[c]{$ u_{2} $}
\Text(5,20)[c]{$ u_{1} $}
\Text(95,70)[c]{$ u_{2} $}
\Text(95,50)[c]{$ d_{2} $}
\Text(55,70)[c]{$ s_{2} $}
\Text(95,20)[c]{$ u_{1} $}
\Text(80,50)[c]{$ W $}
\Text(50,80)[c]{$ \gamma $}
\Line(120,60)(220,60)
\Line(120,40)(220,40)
\Line(120,10)(220,10)
\Photon(190,80)(190,60)3 3
\Photon(160,60)(160,40)3 3
\Text(125,70)[c]{$ s_{1} $}
\Text(175,70)[c]{$ u_{1} $}
\Text(125,50)[c]{$ u_{2} $}
\Text(125,20)[c]{$ u_{1} $}
\Text(215,70)[c]{$ u_{2} $}
\Text(215,50)[c]{$ d_{2} $}
\Text(215,20)[c]{$ u_{1} $}
\Text(150,50)[c]{$ W $}
\Text(180,80)[c]{$ \gamma $}
\Line(240,60)(340,60)
\Line(240,40)(340,40)
\Line(240,10)(340,10)
\Photon(280,40)(280,60)3 3
\Photon(300,20)(300,40)3 3
\Text(245,70)[c]{$ s_{1} $}
\Text(245,50)[c]{$ u_{2} $}
\Text(290,50)[c]{$ d_{1} $}
\Text(245,20)[c]{$ u_{1} $}
\Text(335,70)[c]{$ u_{2} $}
\Text(335,50)[c]{$ d_{2} $}
\Text(335,20)[c]{$ u_{1} $}
\Text(265,50)[c]{$ W $}
\Text(310,30)[c]{$ \gamma $}
\end{picture}
\end{center}
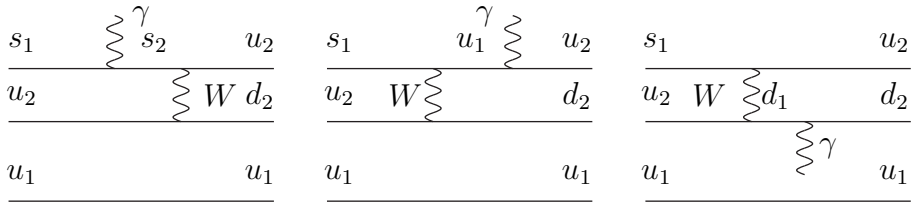
Fig. 5.  The 3-quark diagrams of the decay $ \Sigma^{+}\rightarrow
p \gamma$ corresponding to the matrix element $A_4$ with the 3rd
quark $u_{1}=u_{\uparrow}$ as a spectator.



\begin{thebibliography}{99}
\bibitem{PDG}  Particle Data Group, Journal of Physics G {\bf 33}, 1 (2006).
\bibitem{PDGold} Particle Data Group, The Eur. Phys. J. C {\bf 3},613 (1998);
Phys. Rev. D {\bf 54}, 1-I (1996).
\bibitem{Hara} Y.~Hara, Phys. Rev. Lett. {\bf 12}, 378 (1964).
\bibitem{Sharma} R.C.~Verma and A.~Sharma, Phys. Rev. D {\bf 38}, 1443 (1988).
\bibitem{Verma} A.N.~Kamal and R.C.~Verma, Phys. Rev. D {\bf 26}, 190 (1982).
\bibitem{Ryaz} A.N.~Kamal and Riazuddin, Phys. Rev. D {\bf 28}, 2317 (1983).
\bibitem{ZenM} J.~Lach and P.~Zenczykowski,
Int. J. Mod. Phys. A {\bf 10}, 3817 (1995).
\bibitem{Zen} P.~Zenczykowski,  Phys. Rev. D {\bf 40}, 2290 (1989).
\bibitem{Zen1} P.~Zenczykowski, Phys. Rev. D {\bf 44}, 1485 (1991).
\bibitem{Zen06} P.~Zenczykowski, hep-ph/0512122.
\bibitem{Neu} H.Neufeld, Nucl. Phys.  B {\bf 402}, 166 (1993).
\bibitem{Behr} R.E.~Behrends, Phys. Rev. {\bf 111}, 1691 (1958).
\bibitem{Pak} R.H.~Graham and S.~Pakvasa,
Phys. Rev. 140, B1144 (1965).
\bibitem{Penguin} S.G.~Kamath, Nucl. Phys.  B {\bf 198}, 61 (1982); J.O.~Eeg,
Z. Phys. C {\bf 21}, 253 (1984).
\bibitem{BukGIM} E.N.~Bukina, V.M.~Dubovik, V.S.~Zamiralov,
Nucl. Phys. B (Proc. Suppl.) {\bf 93}, 34 (2001).
\end{thebibliography}
\end{document}